\documentclass[useAMS,usenatbib,usegraphicx,letters]{mn2e}
\usepackage{graphicx, txfonts, color, soul,ulem, multirow}
\usepackage[bookmarks=true]{hyperref}
 \usepackage{subfig}
\usepackage{graphicx}
\usepackage[countmax]{subfloat}

\title[Measuring spot latitude drift rates with transits]{Using \textit{Kepler} transit observations to measure stellar spot belt migration rates}
\author[J. Llama et al.]{J. Llama$^1$\thanks{E-mail: joe.llama@st-andrews.ac.uk}, M.~Jardine$^1$, D.~H.~ Mackay$^2$ and R. Fares$^1$\\
$^1$SUPA, School of Physics \& Astronomy, University of St Andrews, North Haugh, St Andrews,  KY16 9SS, UK\\
$^2$ School of Mathematics \& Statistics, University of St Andrews, North Haugh, St Andrews,  KY16 9SS, UK\\
}

\begin{document}
 
\date{Accepted for publication in MNRAS letters.}

\pagerange{\pageref{firstpage}--\pageref{lastpage}} \pubyear{2011}

\maketitle

\label{firstpage}

\begin{abstract}
Planetary transits provide a unique opportunity to investigate the surface distributions of star spots. Our aim is to determine if, with continuous observation (such as the data that will be provided by the \textit{Kepler} mission), we can in addition measure the rate of drift of the spot belts. 
We begin by simulating magnetic cycles suitable for the Sun and more active stars, incorporating both flux emergence and surface transport. This provides the radial magnetic field distribution on the stellar surface as a function of time. We then model the transit of a planet whose orbital axis is misaligned with the stellar rotation axis. Such a planet could occult spots at a range of latitudes. This allows us to complete the forward modelling of the shape of the transit lightcurve. We then attempt the inverse problem of recovering spot locations from the transit alone. From this we determine if transit lightcurves can be used to measure spot belt locations as a function of time.
We find that for low-activity stars such as the Sun, the 3.5 year \textit{Kepler} window is insufficient to determine this drift rate. For more active stars, it may be difficult to distinguish subtle differences in the nature of flux emergence, such as the degree of overlap of the ``butterfly wings". The rate and direction of drift of the spot belts can however be determined for these stars. This would provide a critical test of dynamo theory.

\end{abstract}

\begin{keywords}
Stars: Activity, Stars: Spots, Stars: Planetary Systems; Planet and Satellites: General;   Sun: Sunspots.
 \vspace{-0.25in}
\end{keywords}

\section{Introduction}

The solar magnetic cycle is now very well observed.  On the Sun, the distribution of solar spots is known to vary over an 11-year cycle. The latitudes at which spots emerge as a function of time gives rise to the well-known ``butterfly diagram". This diagram shows that at the beginning of a solar cycle the emergence of spots preferentially occurs at high latitudes and moves down to lower latitudes towards the end of the cycle.  It was initially assumed that this would be the case on all stars, however, stars such as AB Doradus, a very bright, rapid-rotator have been observed over many years with no such pattern being detected \citep{Donati:1997ui}. By monitoring the variability in stellar brightness over time we can build up a picture of the spot distributions that cause rotational modulations in stellar lightcurves. \cite{Berdyugina:2005wb}  provides a comprehensive review on star spot theory and observations on cool stars.

The transit method of detecting exoplanets has now advanced well beyond the initial detection of hot-Jupiters, with observations revealing more information about both the planet and the host star. Asymmetries in exoplanetary transit lightcurves are proving to be a useful tool in probing both the planetary atmosphere and the interaction between the planet and its host star. \cite{Fossati:2010do} reported an asymmetry in the light curve of WASP-12b when observed in the near-UV. \cite{Vidotto:2010jh} and  \cite{Llama:2011de} have shown that this asymmetry could be caused by a magnetically-channelled stellar wind interacting with the planet's magnetic field. This results in a bow shock transiting ahead of the planet, suggesting that transit observations may be an indirect method for detecting exoplanetary magnetic fields.

Another reported asymmetry in transit lightcurves arises when a planet transits over a star spot, as shown in Figure \ref{fig:cartoon}. When the planet is transiting over a dark region on the stellar disk such as a star spot, the amount of light blocked by the planet becomes lower which produces a positive ``bump" in the lightcurve \citep{Rabus:2009bo}. There have been reports of planets transiting across star spots in many different systems (see for example \cite{Rabus:2009bo,Pont:2007kx,Lanza:2009gd,SanchisOjeda:2011bwa}).  \cite{SilvaValio:2008fw} have shown that detecting star spots through transit observations allows us to determine the longitude of the spot and therefore multiple transit events can help determine the stellar rotation rate. \cite{SanchisOjeda:2011bwa} have also used such transits to investigate the alignment between the stellar rotation axis and the planet's orbital axis by looking for recurring star spot bumps in the transit lightcurves. Whilst most of the work so far has concentrated on individual spots, missions such as \textit{Kepler} provide a unique opportunity to build up a picture of spot evolution and cycles on planet hosting stars by providing highly precise, continuous observations of many transiting systems. 

In this Letter we show that it may be possible to measure the migration rates of stellar activity belts from transit observations such as those provided by \textit{Kepler}. Using simulated solar and stellar cycles we investigate whether this method can distinguish between different types of magnetic cycle.

  \vspace{-0.2in}
\section{The Model}
 \subsection{Simulated butterfly diagrams}
In order to study how transit light curves may help recover the spot evolution on planet-hosting stars, we first specify the emergence of flux through the stellar surface. The choice of the emergence rates and strengths of the bipoles, as well as the time-latitude dependence are important elements to accurately simulate the evolution of star spots \citep{Mackay:2004kja}.

For this investigation we choose various examples of possible solar and stellar butterfly diagrams. Table \ref{table:results} provides an overview of our four models and the left column of Figure \ref{fig:multi_panel} shows the corresponding input flux emergence patterns. Model A is the solar butterfly diagram of \cite{vanBallegooijen:1998jm}. As shown in Figure \ref{fig:solar_butterfly}, this pattern of magnetic flux emergence reproduces the solar 11-year spot cycle. The beginning of the cycle shows the bipoles emerging preferentially at latitudes of $\pm40^\circ$. The bipoles emerge at lower latitudes through the cycle, finally reaching latitudes of $\pm10^\circ$ by the end of the cycle.
 
Models B, C, and D are based on theoretical butterfly diagrams used by \cite{McIvor:2006ja} to simulate magnetic cycles on active stars such as AB Dor. AB Dor is a solar-like rapid rotator ($P_{\rm rot}=0.51$ days), with a stronger magnetic field than the Sun \citep{Donati:1997ui}. Unlike the Sun, which exhibits clear belts of activity whose latitude varies through the magnetic cycle, AB Dor shows spots at all latitudes at each observing epoch  \citep{Jeffers:2007bi}. The flux emergence patterns used by \cite{McIvor:2006ja}  in conjunction with a model for the transport of this flux across the stellar surface reproduce this observed behaviour. For ease of comparison with the Sun, \cite{McIvor:2006ja} assume an 11-year cycle in their simulated butterfly diagrams, which will also be adopted here.  Model B, shown in Figure \ref{fig:abdor_solar_butterfly}, consists of an enhanced butterfly pattern, with the maximum latitude of emergence extended to $70^\circ$. Model C (Figure \ref{fig:abdor_overlapped_butterfly}), is an overlapping butterfly pattern, again with the maximum latitude of emergence increased. This input allows flux to appear at both high and low latitudes at all times but also retains the butterfly pattern. Model D (Figure \ref{fig:abdor_no_butterfly}), assumes no latitude-time relationship and the emerging bipoles are completely random in both longitude and latitude throughout the cycle.

\begin{figure}    \centering
   \includegraphics[width=2in]{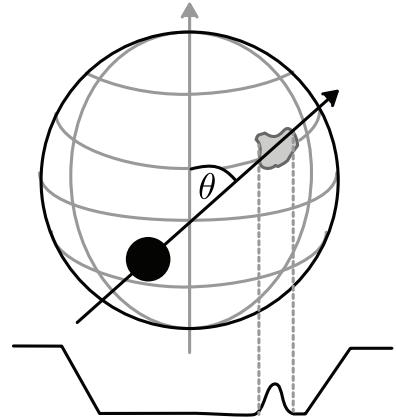} 
   \caption{Schematic of the model (not drawn to scale). The planet transits the star on an inclined orbit.  The angle $\beta$ defines the obliquity of the planet. When the planet transits over a dark region of the stellar disk such as a star spot, a bump appears in the transit lightcurve.}
   \label{fig:cartoon}
   \vspace{-0.1in}
\end{figure}
\vspace{-0.2in}
\subsection{Surface evolution}

\begin{figure*}
\vspace{-0.75in}
	\centering
	\subfloat[]{
		\label{fig:solar_butterfly}
		\includegraphics[width=0.4\textwidth]{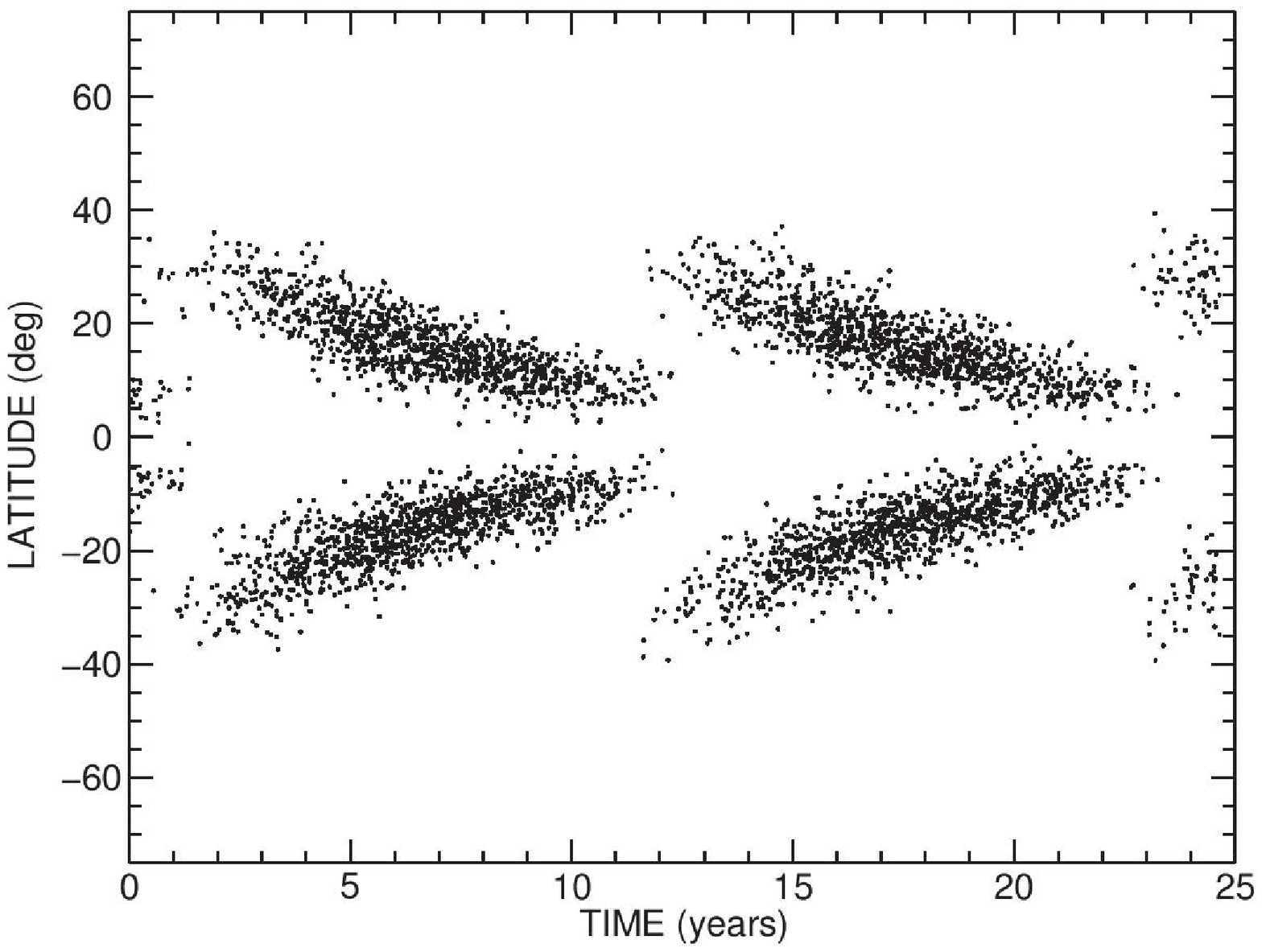}}
	\subfloat[] {
		\label{fig:solar_results}
			\includegraphics[width=0.4\textwidth]{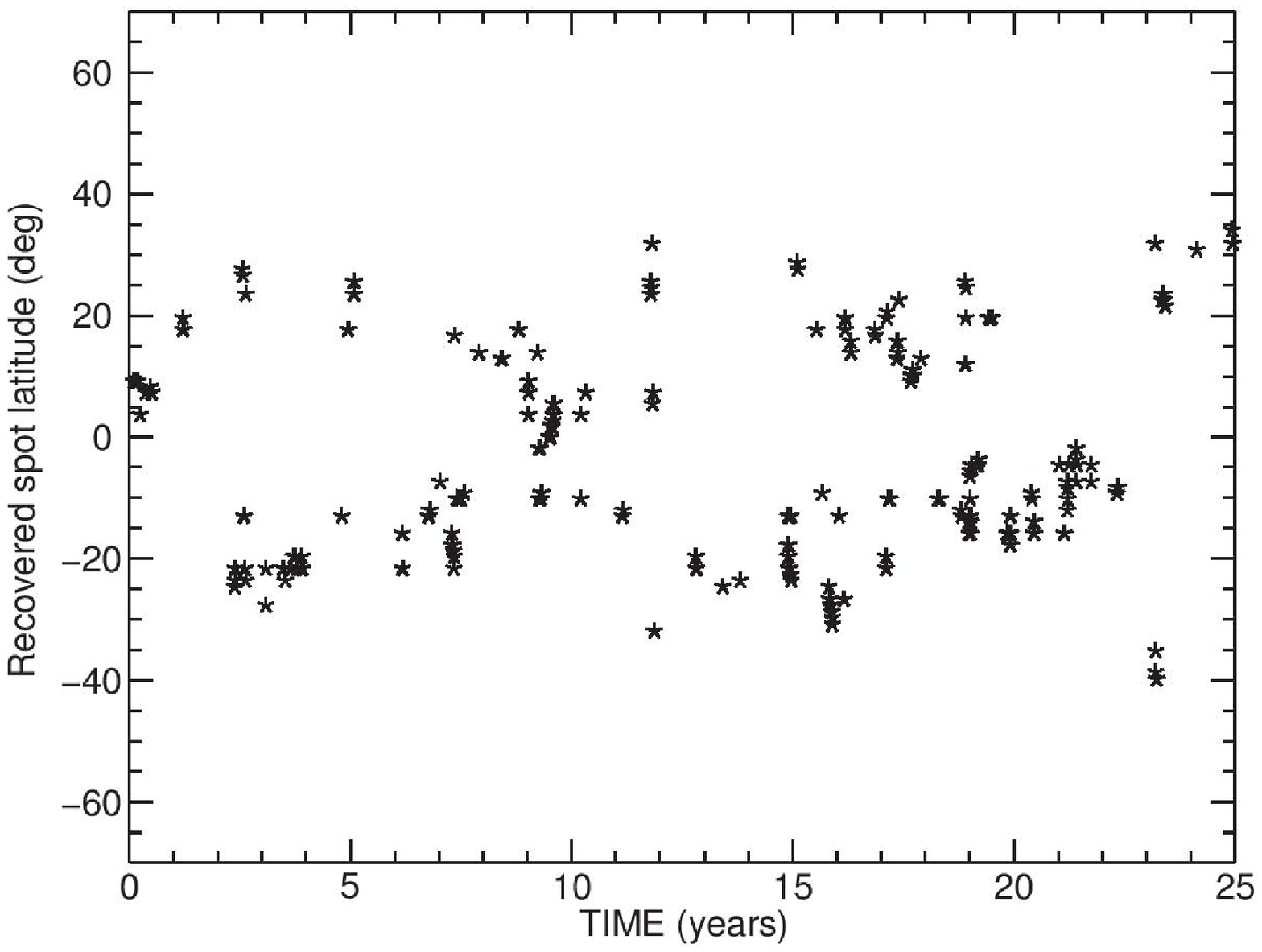}} \\
	\vspace{-0.18in}	
	\subfloat[] {
		\label{fig:abdor_solar_butterfly}
		\includegraphics[width=0.4\textwidth]{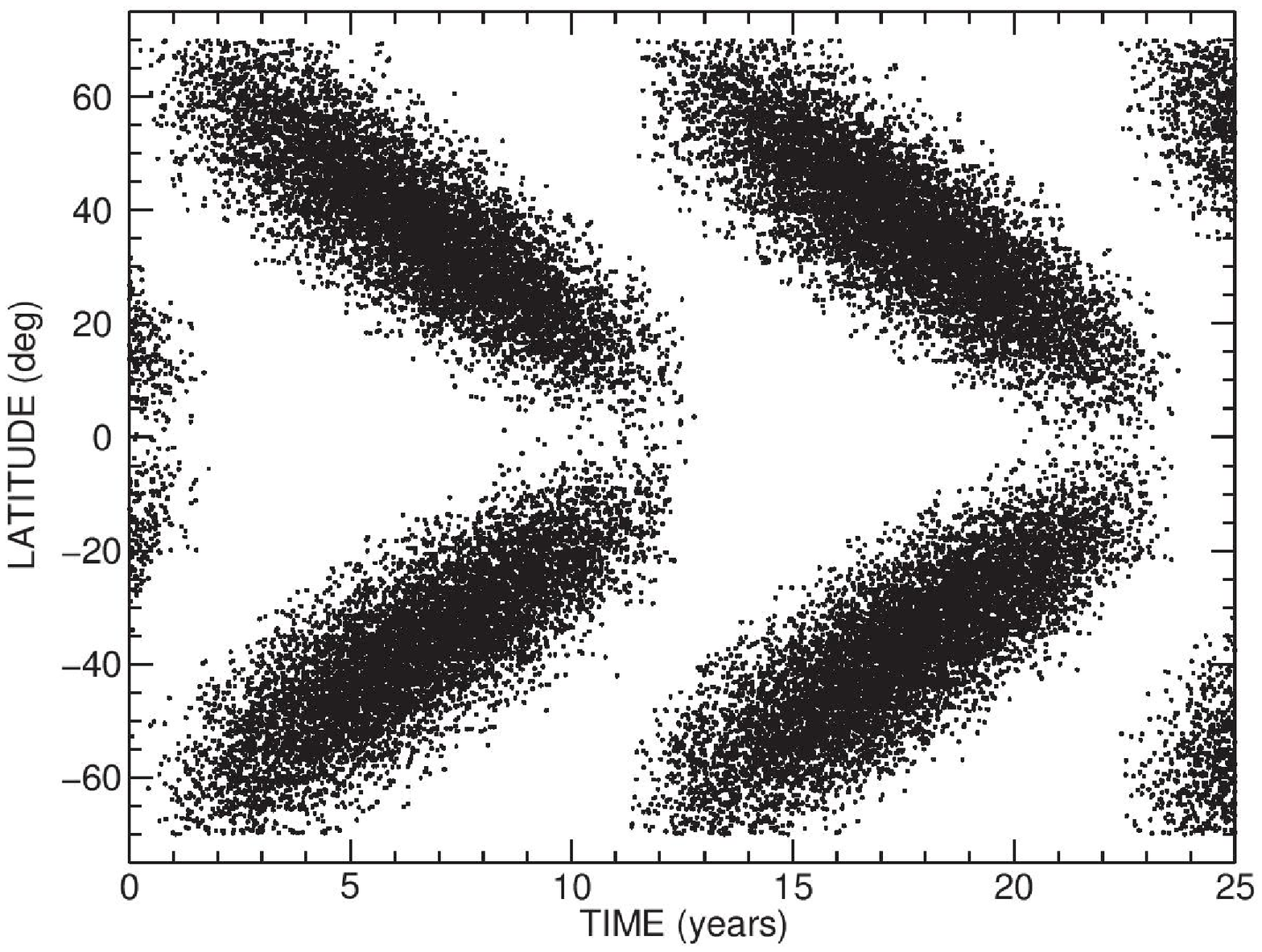}}
	\subfloat[] {
		\label{fig:abdor_solar_results}
		\includegraphics[width=0.4\textwidth]{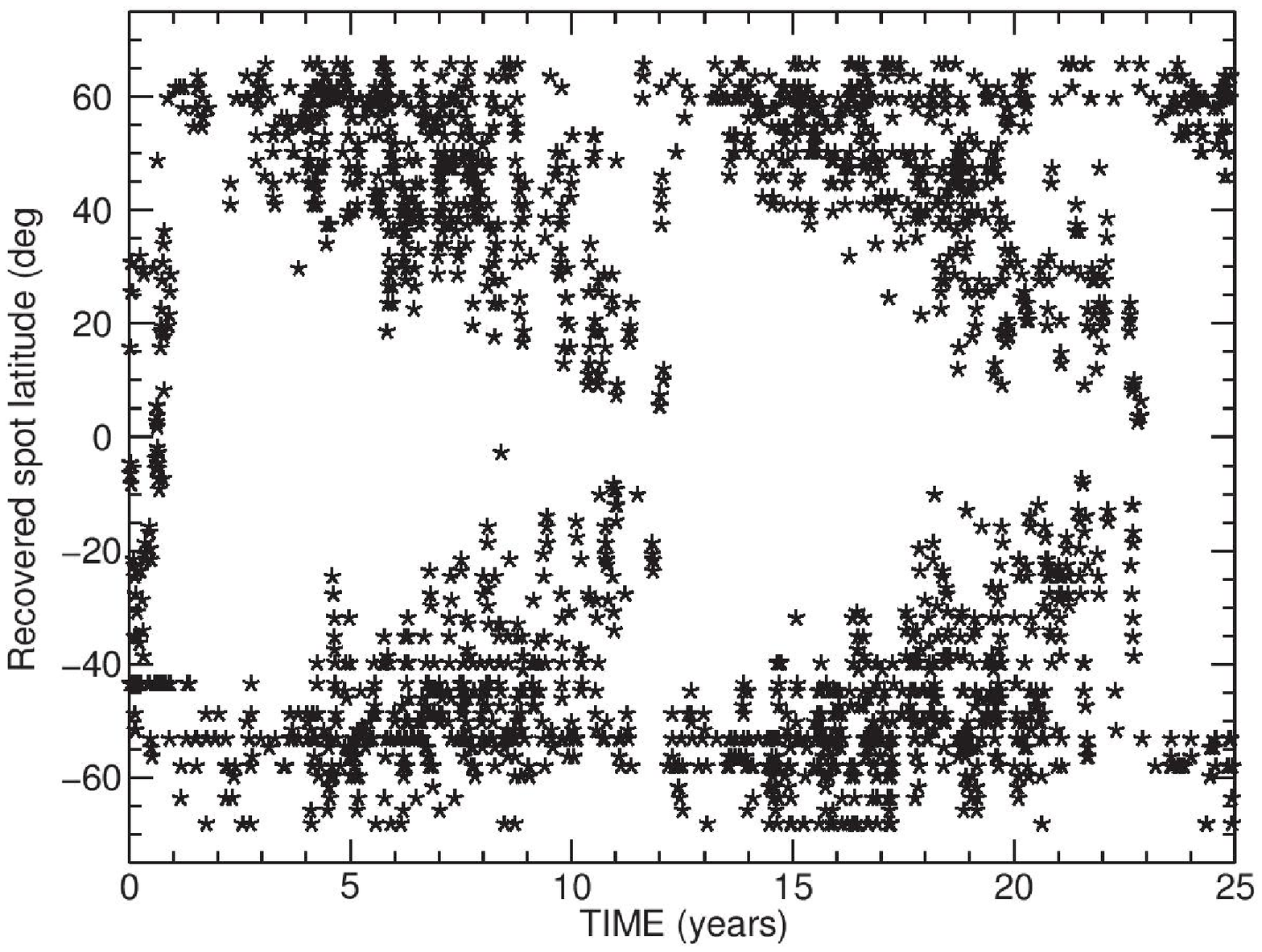}	} \\
		\vspace{-0.18in}	
	\subfloat[] {
		\label{fig:abdor_overlapped_butterfly}
		\includegraphics[width=0.4\textwidth]{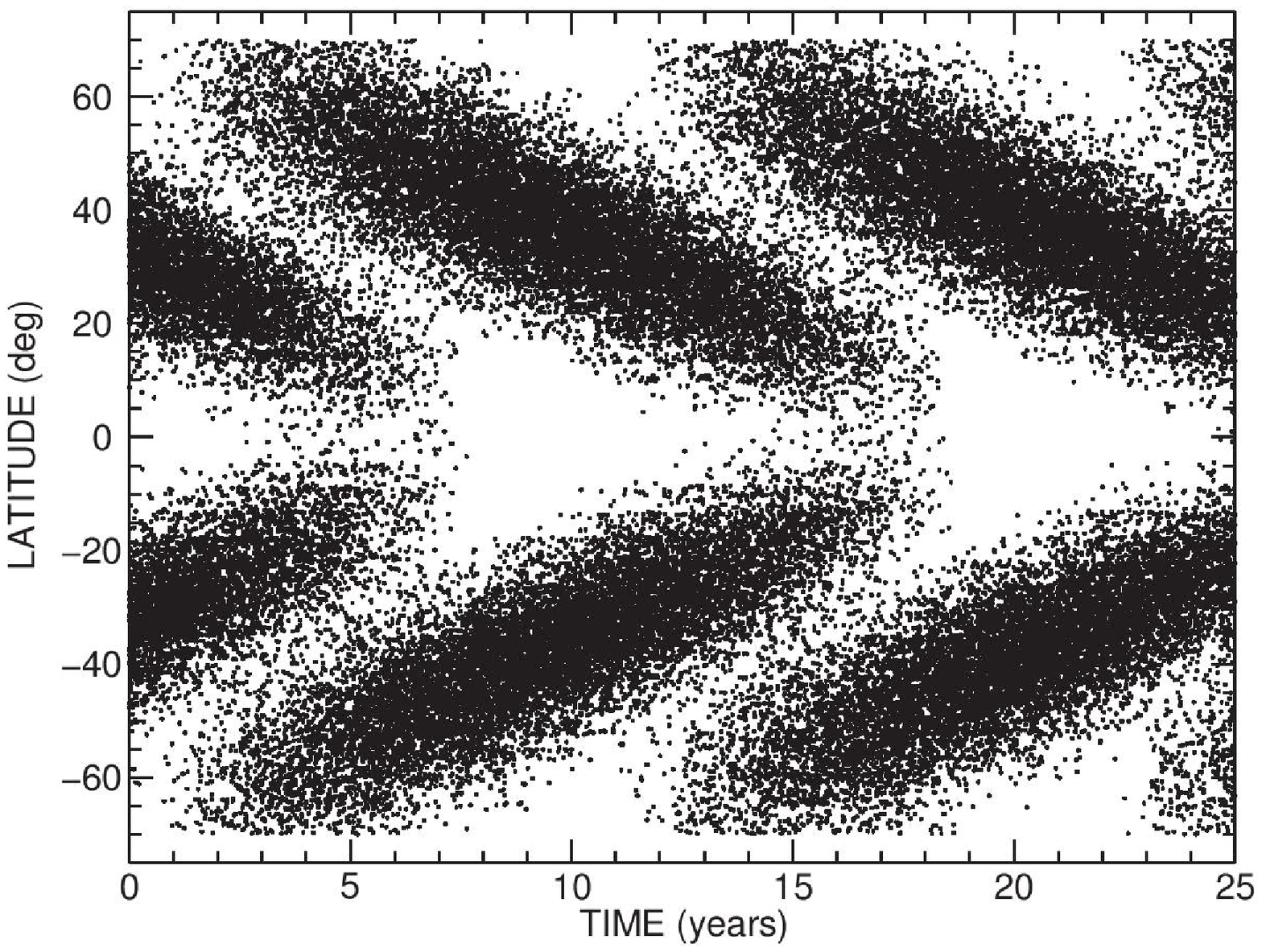}}
	\subfloat[] {
		\label{fig:abdor_overlapped_results}
		\includegraphics[width=0.4\textwidth]{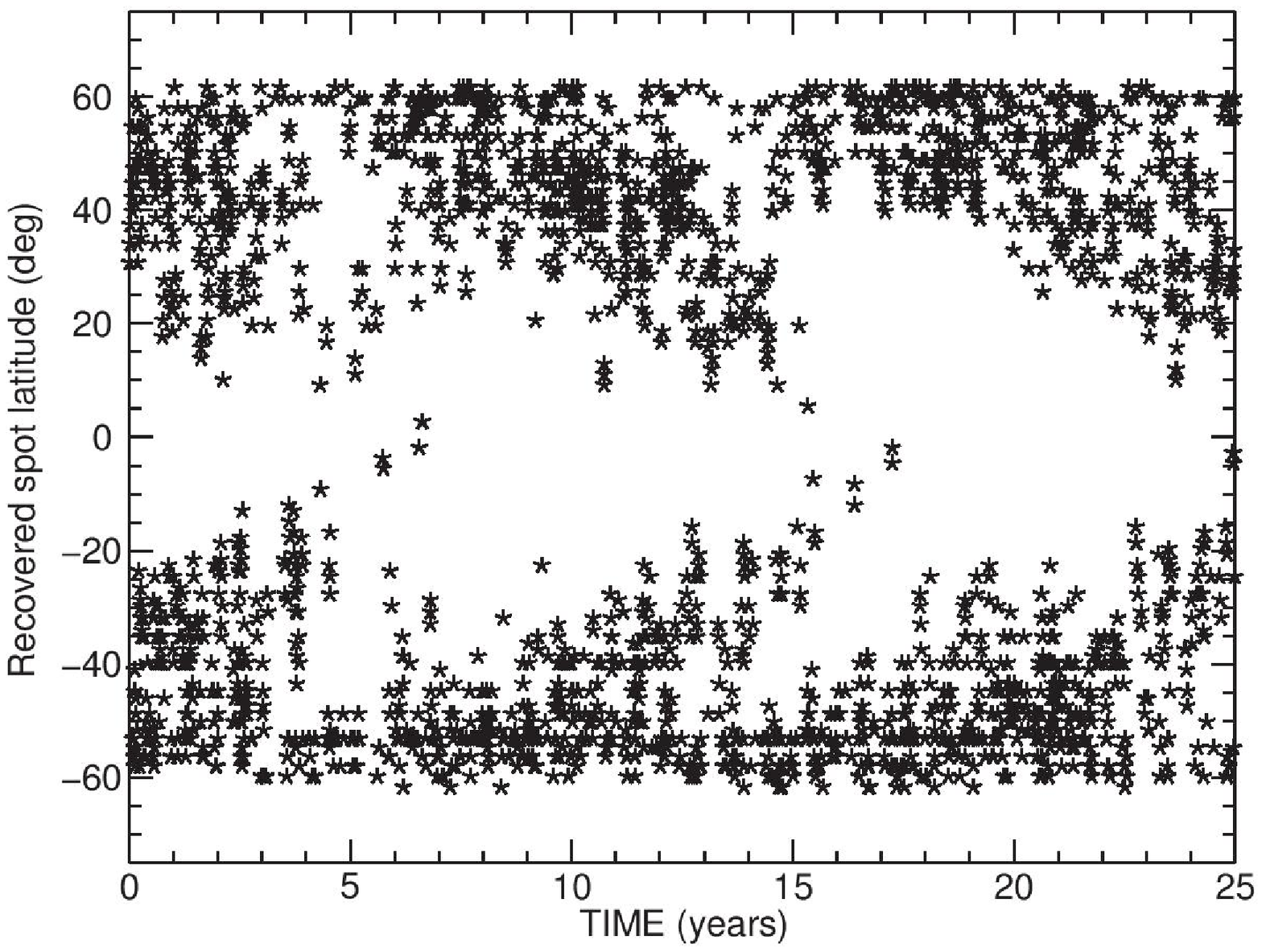}} \\
\vspace{-0.18in}	
	\subfloat[] {
		\label{fig:abdor_no_butterfly}
		\includegraphics[width=0.4\textwidth]{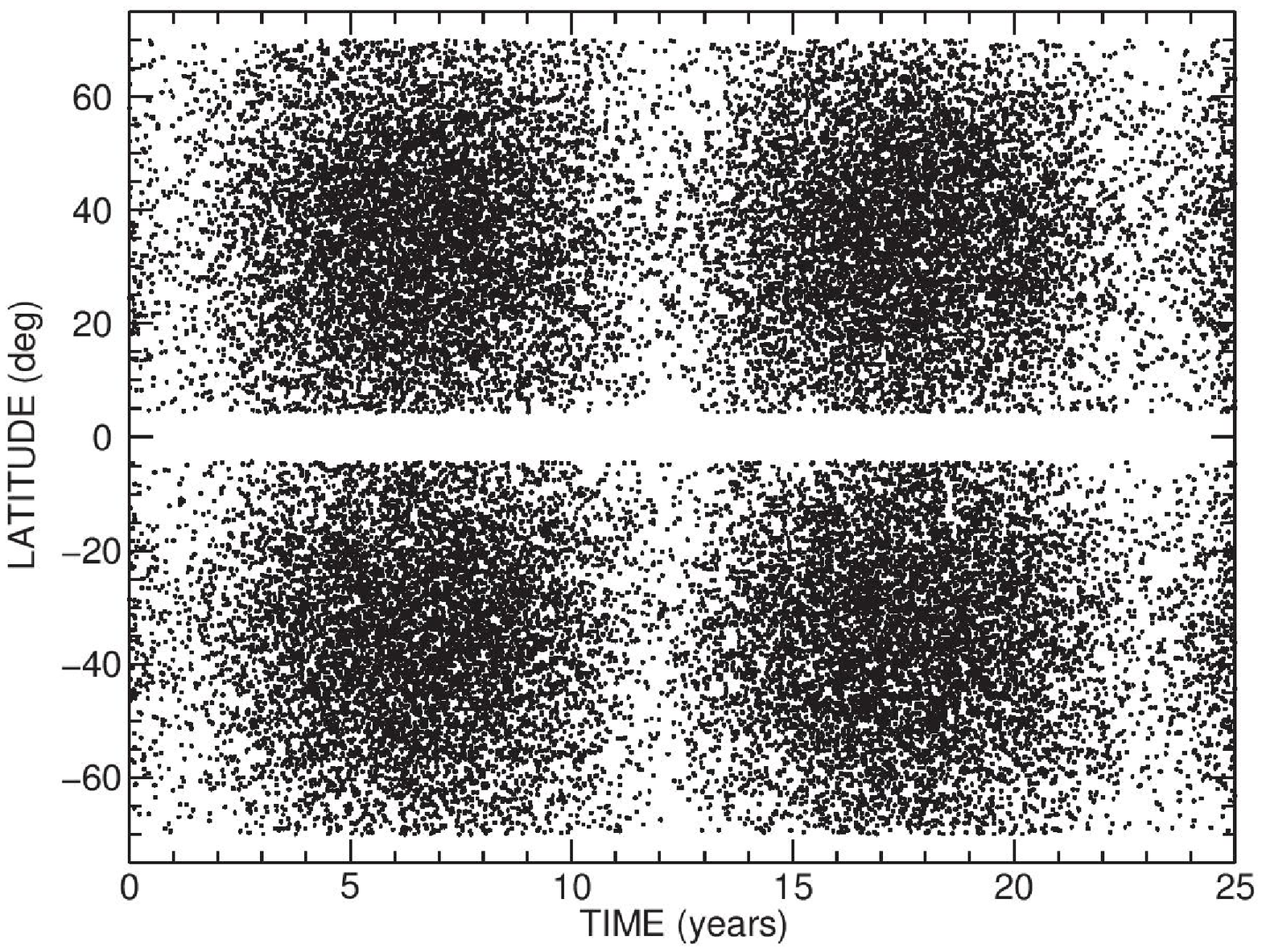}	}
	\subfloat[] {
		\label{fig:abdor_no_results}
		\includegraphics[width=0.4\textwidth]{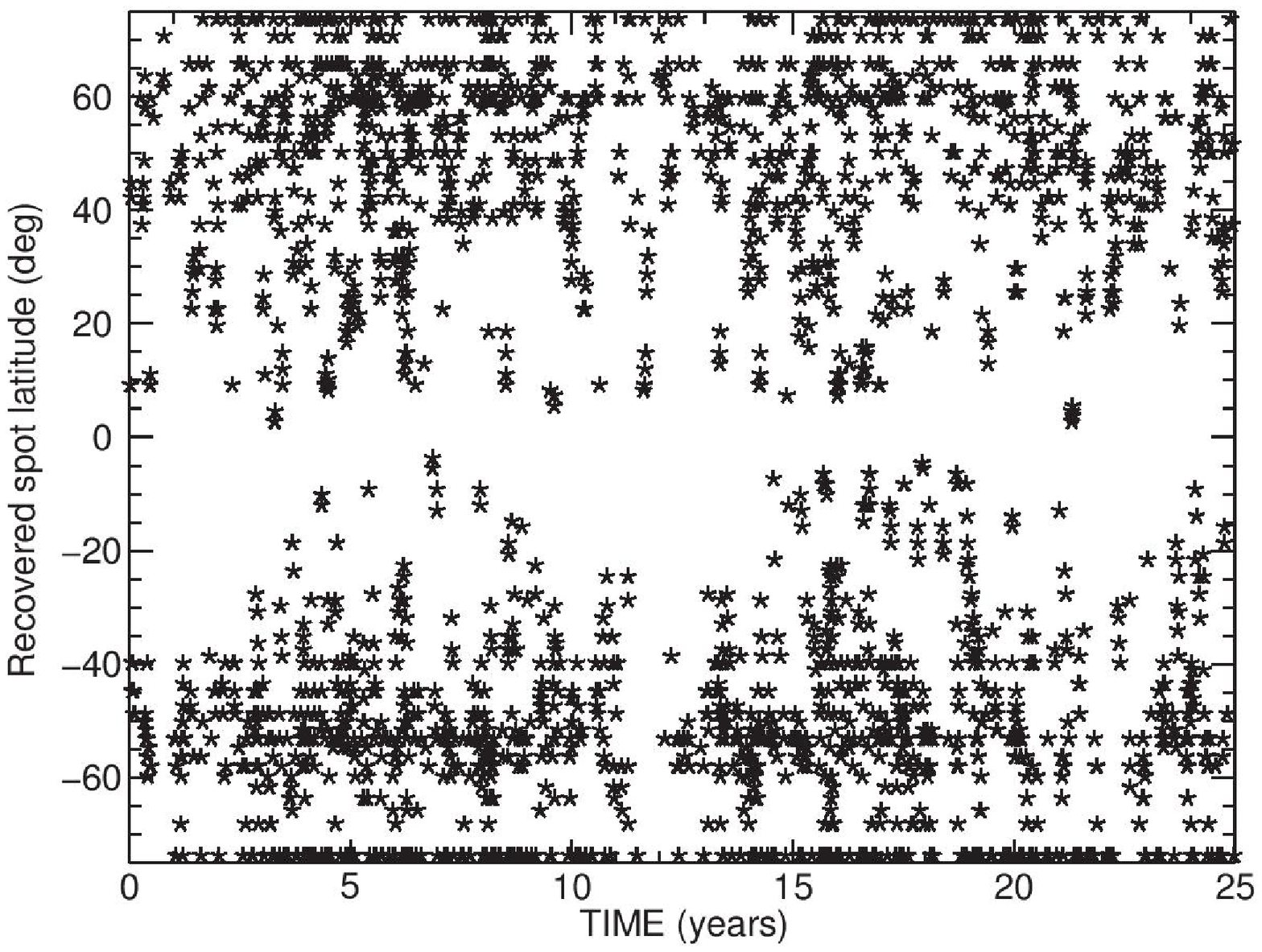}}	
		\vspace{-0.02in}
	\caption{\label{fig:multi_panel} The left column shows the input flux emergence patterns for each of our models \citep{vanBallegooijen:1998jm,McIvor:2006ja}. The right column shows the distribution of spots recovered through the transit method as explained in Section 2.3. The top row corresponds to Model A, a Solar type star. The final three rows correspond to Models B, C, and D which represent the rapid-rotator with an enhanced butterfly pattern, an overlapping butterfly pattern and, no butterfly pattern respectively. }
\end{figure*}

Once the magnetic flux has emerged through the stellar surface it will be transported across the stellar surface by differential rotation, meridional flow and diffusion. It is the coupling of the flux emergence with these transport processes that results in a cyclic behaviour of the surface flux. To reproduce the solar and stellar cycles we use a magnetic flux transport code as described fully in \cite{Mackay:2004kja} (and references therein). By accounting for the effects of differential rotation, meridional flow and super granular diffusion, we evolve the stellar magnetic field at the surface of the star, i.e. $r=R_\star$. The model only accounts for the large-scale magnetic field with spatial scales $>30$Mm, but has been shown to accurately reproduce the solar activity cycle (see for example, \cite{Wang:1991bp}).  The simulated butterfly diagrams (Figures \ref{fig:solar_butterfly}, \ref{fig:abdor_solar_butterfly}, \ref{fig:abdor_overlapped_butterfly} and \ref{fig:abdor_no_butterfly}) are taken as input to the code which evolves the radial magnetic field $B_r(R_\star,\theta,\phi,t)$ by solving the partial differential equation:
\begin{eqnarray}\label{eqn:evolution}
\nonumber \frac{\partial B_r}{\partial t}=\frac{1}{\sin\theta} \frac{\partial}{\partial\theta}\left[\sin\theta\left(-u(\theta)B_r+D\frac{\partial B_r}{\partial\theta}\right)\right] + \frac{D}{\sin^2\theta}\frac{\partial^2B_r}{\partial\phi^2}-\Omega(\theta)\frac{\partial B_r}{\partial\phi},
\end{eqnarray}
where $D$ is the photospheric diffusion constant, $\Omega(\theta)$ is the profile for differential rotation and $u(\theta)$ describes the meridional flow. The parameters used for each model are given in Table \ref{table:results}.

No direct measurements of meridional flow have been made on main sequence stars. We therefore assume a solar profile, where the flow is directed poleward and is expressed in terms of latitude $(\lambda = \pi/2-\theta)$ by the function:
\begin{equation}
\label{eqn:mflow}
	u(\lambda) = \left\{ 
 				 \begin{array}{l l}
				    -u_0\sin(\pi\lambda/\lambda_0), &\quad |\lambda|<\lambda_0 \\
				   0,&\quad \textrm{otherwise}. \\
			  	\end{array} 
				\right.
 \end{equation}
Under this prescription, the meridional flow velocity vanishes above $\lambda_0$. We use the observed values for the Sun of $\lambda_0 =75^\circ$ and $u_0=11 \textrm{ ms}^{-1}$ \citep{Snodgrass:1996bi,Hathaway:1996ee} for the solar type star and $u_0=100 \textrm{ ms}^{-1}$ for the rapid rotator \citep{Mackay:2004kja}. 
For all models, the differential rotation is expressed using the solar profile of \citep{Snodgrass:1983cr},
\begin{equation}
\label{eqn:diffrot}
	\Omega(\theta) = 13.38 - 2.30\cos^2\theta - 1.62\cos^4\theta - \Omega_0 \textrm{ deg day}^{-1}
\end{equation}
 where $\Omega_0$ is the Carrington rate (13.20 deg d$^{-1}$). 
 Finally, the super granular diffusion constant $D$, determines the timescale on which flux moves from the centre of granular cells to the boundaries where it can interact and cancel with flux of opposite polarity. This is given the solar value of $D=450$ km$^2$ s$^{-1}$ \citep{Leighton:1964bv}.  The evolution code outputs a radial magnetic map at each time step, which is set here to be three days to simulate the typical times between a transit event. 

\begin{table}

\centering
	\caption{Parameters used for each of our models. For each model $R_\star=R_\odot$, $R_p=R_J$, and $\beta =30^\circ$ (see Figure \ref{fig:cartoon}). The second column gives the type of input butterfly pattern used; Column three is the initial meridional flow speed; Column four is the diffusion constant; Column five is the drift rate of the input latitudes of flux emergence; Column six is the drift rate of  the predicted spot latitudes.}
\vspace{-0.1in}
 \begin{tabular}{@{}c@{\,\,}c@{\,\,}c@{\,\,}c@{ \,\,}c@{\,\, }c@{\,\, }c@{ }c@{}}
	\hline
	Model  & Type & $u_0$ & $D$ &$P_{\rm rot}$&  Input drift & Recovered drift \\
	& & $(\textrm{ms}^{-1})$ & $(\textrm{km}^{2}\textrm{ s}^{-1})$   &(days)&(deg yr$^{-1}$)& (deg yr$^{-1}$)\\
	\hline
	\hline
	A & Solar & 11 & 450 &27& -2.2 & -2.0  \\
	&butterfly &&&&&\\
	\hline
	B& Rapid rotator & 100 & 450 &0.5& -4.8 & -4.5  \\
	&butterfly &&&&&\\
	\hline
	C& Rapid rotator & 100 & 450 &0.5& -3.3 & -2.5  \\
	&overlapped &&&&&\\
	\hline
	D& Rapid rotator & 100 & 450 &0.5&0 & 0  \\
	&no butterfly &&&&&\\
	\hline
	\end{tabular} 
	\label{table:results}
\vspace{-0.2in}
\end{table}
\vspace{-0.2in}
\subsection{Simulated lightcurves}
To produce the lightcurves we assume a star of radius $R_\star=R_\odot$ and a hot-Jupiter of radius $R_p = R_J$,  so that a 1\% dip in flux will be recorded as the planet transits across the stellar disk (assuming the portion of the stellar disk occulted by the planet has intensity 0).  A schematic of the setup is shown in Figure \ref{fig:cartoon}. We take the impact parameter, $b=0$ so that the planet transits across the centre of the stellar disk. The angle between the stellar rotation axis and the orbital plane of the planet (as shown in Figure \ref{fig:cartoon}) is taken to be, $\beta = 30^\circ$.  Under this setup, the planet transits over a large range of latitudes on the stellar disk and also spends an equal amount of time in each hemisphere. If the planet were to spend more time transiting over one hemisphere of the stellar disk an observational bias may be present in the recovered distribution of spots.  We also assume that $P_{\rm orb} = 3$ days. We set the stellar rotation period to be $P_{\rm rot}=27$ days for the solar type star and $P_{\rm rot} = 0.5$ days for the rapid rotator. In all cases the star is edge on to the observer.

At every transit crossing event (i.e. once every three days) our surface flux transport simulations output the magnetic configuration on the surface of the star allowing for both the rotation of the star and also foreshortening. From this we classify regions of strongest magnetic field ($\rm{|B|}>0.9\,\rm{|B|}_{\rm max}$) as spots.  

We irradiate the stellar disk with Monte Carlo photon packets to reproduce the spatial intensity distribution of a limb-darkened star. We use the limb-darkening law as given by \cite{Claret:2004jy}, where the intensity $I$ is given by 
\begin{equation}
	\frac{I(\mu)}{I(0)} = 1-\sum^4_{n=1}a_n(1-\mu^{n/2}),
	\label{eqn:limb_darkening}
\end{equation}
where $\mu = \cos\theta = (1-r^2)^{1/2}$, $0\le r\le 1$. The coefficients $a_n$ are chosen to match the limb darkening of \textit{Kepler} observations \citep{Sing:2010he}. For this analysis, we choose the limb-darkening coefficients for a star with T$_{\rm eff}=7500$ K and $\log g=4.5$. 
The spatial intensity distribution is also affected by the spot distribution. For simplicity, we assume that regions that contain a spot have an intensity of 0, i.e. are completely dark. We then sum the total radiated flux and normalise by the total number of photons to create the transit lightcurve. 

Any instances where the planet occults a spot, or part of a spot then the fractional loss of light will be less and a positive bump will be registered in the lightcurve. Should the planet occult multiple spots in the same transit we are able to recover each spot crossing separately. The phase of any bumps can then be converted back into latitude on the star. The presence of noise in the data of course limits the sizes of spots that can be recovered. Guided by the results of \cite{SanchisOjeda:2011hd} we only recover spots where the deviation in normalised flux is larger than $0.001\%$. 
 The phase of the bump provides the position of the planet $(x,y)$ on the projected stellar disk. This location is then converted into latitude on the star using using the equation:

\begin{equation}
\lambda = \sin^{-1}\left[\frac{y\sin\left(\sqrt{x^2+y^2}\right)}{\sqrt{x^2+y^2}}\right].
\label{eqn:latitude}
\end{equation}
 By plotting spot latitudes as a function of time, we can therefore determine the rate and direction of drift of the spot belts.

  \begin{figure*}    \centering
      \includegraphics[width=0.33\textwidth]{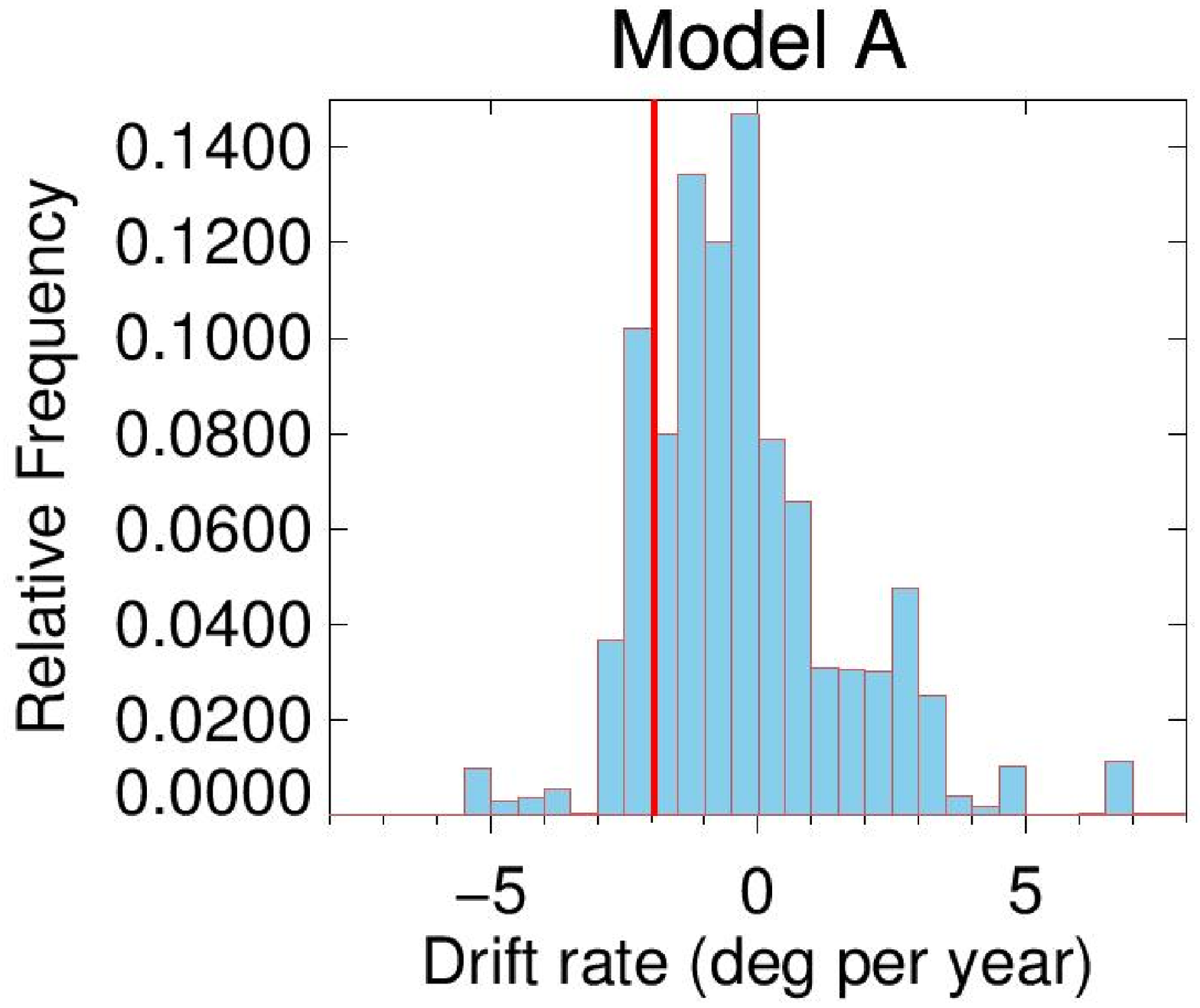}
   \includegraphics[width=0.33\textwidth]{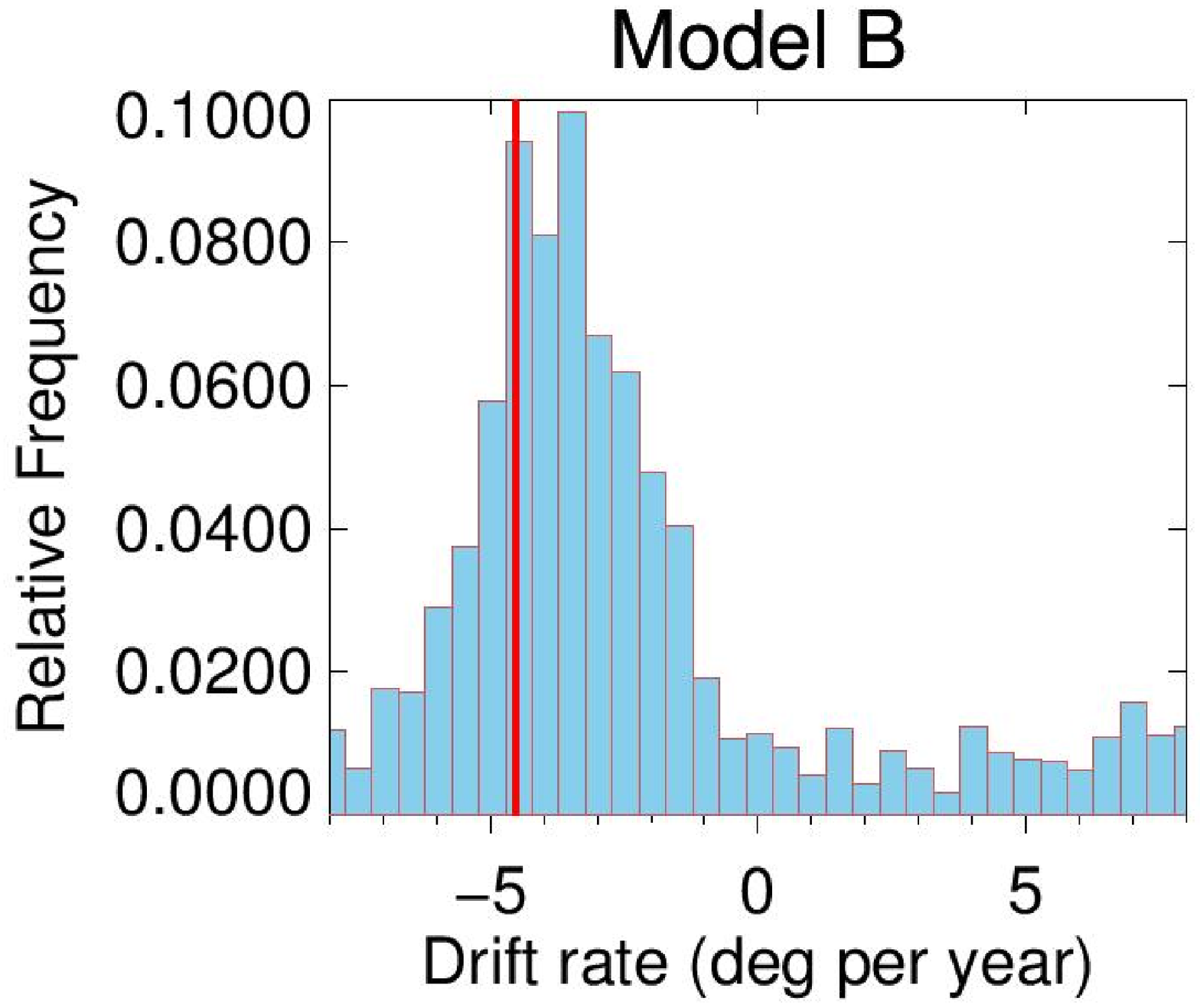}
   \includegraphics[width=0.33\textwidth ]{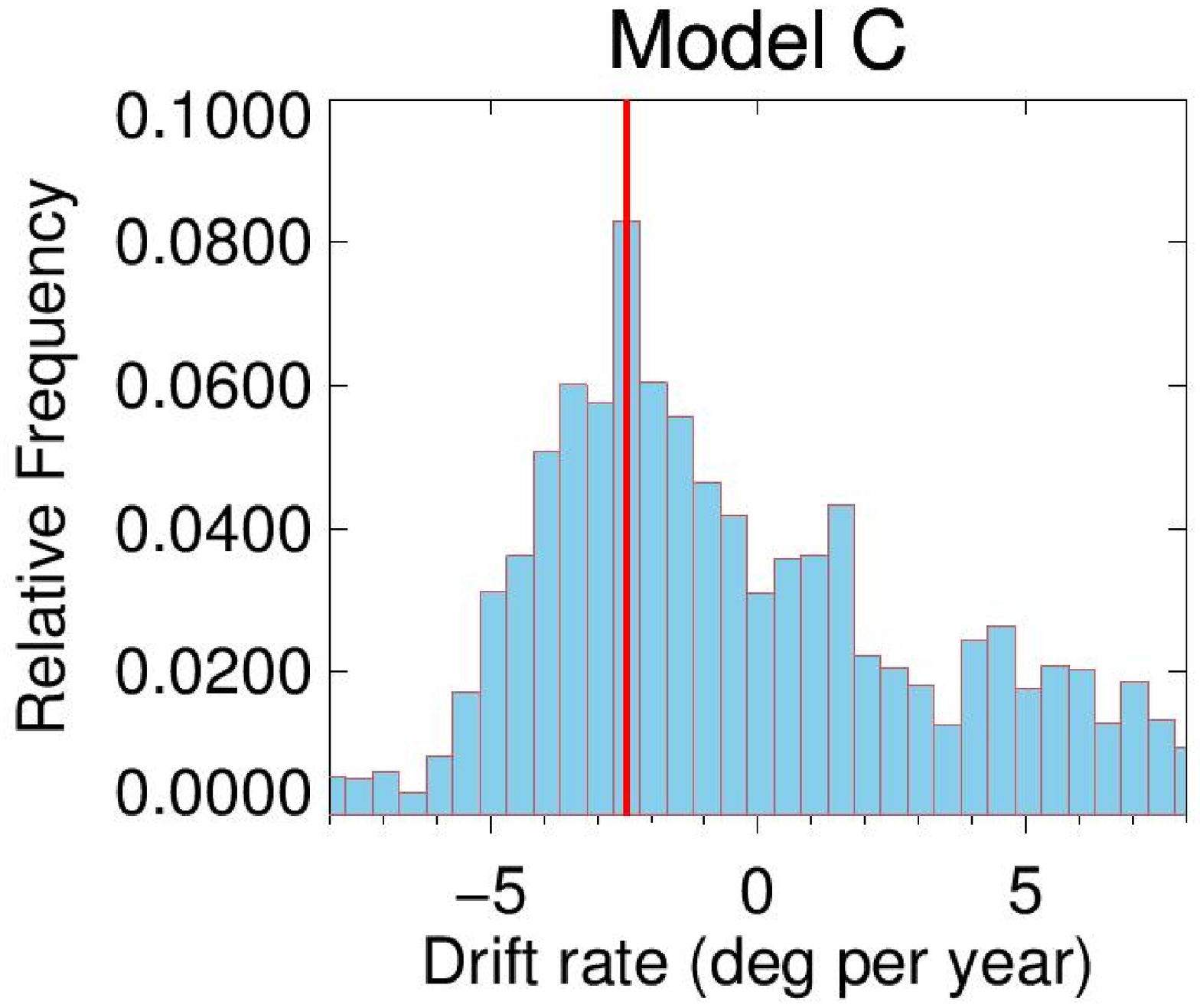}
  
   \caption{Histograms comparing the drift rates of spot belts calculated from (a) the full simulation (red-line) with (b) the values that can be recovered from taking 10,000 3.5-year ``snapshots" of the data which simulates the current \textit{Kepler} observing window. Table \ref{table:results} summarises the parameters used in each model.}
      \label{fig:histogram}
\vspace{-0.2in}
\end{figure*}

\vspace{-0.25in}
\section{Results}
 The results of our simulations  are shown in the right hand column of Figure \ref{fig:multi_panel}. The recovered spot locations for Model A, a Solar type cycle are shown in Figure \ref{fig:solar_results}. The results indeed show a time-latitude relation similar to the input flux emergence diagram. The recovered spots are at higher latitudes at the beginning of the cycle than at its end, with the pattern repeating after 11-years. 
 
The recovered spots of the rapid rotator are just as revealing. Figure \ref{fig:abdor_solar_results} shows the results for Model B, the enhanced  butterfly diagram (without an overlap). The recovered spots show a cycle repeating on an 11-year scale which matches the input.  The spots appear at high latitudes at the beginning of the cycle and lower latitude spots appear towards the end of the cycle. Although  the input pattern is reproduced,  we notice high-latitude spots recovered throughout the cycle. This is to be expected due to the increased rotation rate of the star and also the increased meridional flow (compared to the solar values) that drives the magnetic field poleward \citep{McIvor:2006ja}. Even in the phase of the cycle where spots emerge at low latitude, they are rapidly pushed to higher latitudes because of the meriodional flow, which results in high-latitude spots being more readily recovered throughout the cycle. 

Figure \ref{fig:abdor_overlapped_results} show the resultant spot latitudes for the overlapped butterfly diagram (Model C). This again shows that it is possible to recover a time-latitude dependence but suggests that it may   be difficult to distinguish it from Model B (Figure \ref{fig:abdor_solar_results}).  
 
Finally, Figure \ref{fig:abdor_no_results} shows the recovered spot latitudes for a cycle with no time-latitude dependence (Model D). Although the input diagram has no time-latitude dependence, the results show a weak latitude dependence, with more spots appearing at higher latitudes. This is likely due to the increased Coriolis force and meridional flow pushing the spots poleward.  One interesting result is that the recovered spots do not appear above $60^\circ$ even though \citep{McIvor:2006ja} report magnetic activity at latitudes up to the pole. 

For this method to recover spots at very high latitudes, the planet must be on an highly inclined orbit (i.e. the value of $\beta$ in Figure \ref{fig:cartoon} must be very small). Foreshortening of the spots on the stellar disk causes high latitude spots to appear smaller than similar spots at lower latitude. This method is therefore less sensitive to detecting spots at high latitude and preferentially finds spots at lower latitudes than spots near the pole.

To further compare the results of our simulations with the input, we calculate the drift rate of both the input spot emergence pattern through the cycle and the drift rate of the recovered spots. For the input flux emergence we are able to calculate the drift rate directly. However, since the recovered spots have been subjected to the surface evolution processes described in Section 2.2, the drift rate is recovered by measuring the slope of the lower envelope of each butterfly wing. This is achieved by taking 0.1-year time bins, and then calculating the lower 90$^{th}$ percentile of each bin. We then fit the function $\lambda(t)=A+B t$, where $A$ is the initial latitude and $B$ is the drift rate. The recovered value of the drift rates are shown in the final two columns of Table \ref{table:results}. In all four models the values compare favourably, however, the recovered drift rate is lower than the input rate. This is likely due to the meridional flow moving flux poleward. 
 
\vspace{-0.2in}
\section{Discussion} 
 	
In this Letter we have addressed the possibility of recovering spot cycles on stars hosting a transiting planet. This study is particularly timely because of the availability of  the \textit{Kepler} data. By modelling the emergence and transport of flux on the surface of  a star throughout its magnetic cycle, we have predicted the distribution of spots as a function of time. From this forward-modelling we have demonstrated that transit lightcurves can be used to recover the drift rates of spot latitudes on active stars. This rate of drift of spot latitudes is a critical prediction of dynamo theories, but it has in the past been particularly difficult to observe directly. While other methods such as Doppler imaging can provide a snapshot of the spot distribution, only continuous viewing capability such as that of  \textit{Kepler} can provide a sufficiently long time series to measure drift rates. The recovered drift rates of the active latitudes reveal the nature of the dynamo - whether it is a solar-like behaviour where the active latitudes drift towards the equator over the cycle, or anti-solar where they drift towards the poles.

We have used various simulated spot cycles, assuming a cycle length of 11-years, similar to that of the Sun. \textit{Kepler} is currently scheduled to collect data for 3.5 years. To investigate whether it is possible to detect a drift in the recovered spot positions over this time we have randomly   	placed 3.5 year windows over each of our result sets. We then carry out the analysis as described in the final paragraph of Section 3 to determine the drift rate for each window. By repeating this 10,000 times we have investigated whether a consistent value for the drift rate can be recovered. 

Figure \ref{fig:histogram} shows the results for the data in the Northern hemisphere.  We have carried out this analysis on Models A, B, and C (top three rows of Figure \ref{fig:multi_panel}). Model D  (the bottom row of Figure \ref{fig:multi_panel}) is excluded from this analysis because it was designed to not exhibit any time-latitude dependence.
  
For the Solar cycle (Model A) we find that a 3.5-year window of observation is not long enough to reveal a time-latitude dependence in the data. The solar case is hampered by the relatively low activity rate of the star, and hence, the likelihood of a spot being discovered through transit methods is significantly lower than for a more active star. To determine a time-latitude dependence for a solar type star a longer set of observations would likely be required. 

For more active stellar cycles (Models B, and C) the 3.5-year window produces a stronger Gaussian shaped distribution suggesting it is possible to recover a time-latitude dependence in the data. 
We find however that even if the general trend of spot locations and evolution can be recovered, it may be difficult to distinguish between an enhanced spot cycle and an enhanced but overlapped spot cycle. 

The method of using transiting planets to determine changes in spot latitudes is particularly powerful, but requires not only that a planet is transiting, but also on an inclined orbit. Low latitude spots are preferentially recovered by this method, and shorter magnetic cycles will be easier to study within the \textit{Kepler} lifetime.

We have shown that the data collected using the observational window of the \textit{Kepler} space mission has the potential to determine the drift rates of spot belts on active stars. Transit observations can therefore provide new insight into stellar activity and cycles, and hence are an additional, critical test for stellar dynamo theories.
 
\vspace{-0.27in}
\section*{Acknowledgments}
JL acknowledges the support of an STFC studentship. We would like to thank K. Wood,  A.C. Cameron for helpful discussions and advice and also A.A. van Ballegooijen who wrote the original flux transport codes used in this letter.
\vspace{-0.27in}
 \bibliography{spots} 

\label{lastpage}

\end{document}